\documentclass[fleqn,10pt]{wlscirep}
\title{Indication of band flattening at the Fermi level in a strongly correlated electron system}

\author[1]{M.~Yu. Melnikov}
\author[1]{A.~A. Shashkin}
\author[1]{V.~T. Dolgopolov}
\author[2,3]{S.-H. Huang}
\author[2,3]{C.~W. Liu}
\author[4,*]{S.~V. Kravchenko}
\affil[1]{Institute of Solid State Physics, Chernogolovka, Moscow District 142432, Russia}
\affil[2]{Department of Electrical Engineering and Graduate Institute of Electronics Engineering, National Taiwan University, Taipei 106, Taiwan}
\affil[3]{National Nano Device Laboratories, Hsinchu 300, Taiwan}
\affil[4]{Physics Department, Northeastern University, Boston, Massachusetts 02115, USA}

\affil[*]{s.kravchenko@northeastern.edu}

\begin{abstract}
Using ultra-high quality SiGe/Si/SiGe quantum wells at millikelvin temperatures, we experimentally compare the energy-averaged effective mass, $m$, with that at the Fermi level, $m_F$, and verify that the behaviours of these measured values are qualitatively different. With decreasing electron density (or increasing interaction strength), the mass at the Fermi level monotonically increases in the entire range of electron densities, while the energy-averaged mass saturates at low densities. The qualitatively different behaviour reveals a precursor to the interaction-induced single-particle spectrum flattening at the Fermi level in this electron system.
\end{abstract}
\begin{document}

\flushbottom
\maketitle

\thispagestyle{empty}

The creation and investigation of flat-band materials is a forefront area of modern physics.\cite{heikkila2011,ns2013,peotta2015,volovik15} The interest is ignited, in particular, by the fact that, due to the anomalous density of states, the flattening of the band may be important for the construction of room temperature superconductivity. The appearance of a flat band is theoretically predicted \cite{amusia14,camjayi08,yudin14} in a number of systems including heavy fermions, high-temperature superconducting materials, $^3$He, and two-dimensional (2D) electron systems. As the strength of fermion-fermion interaction is increased, the single-particle spectrum becomes progressively flatter in the vicinity of the Fermi energy eventually forming a plateau. The flattening of the spectrum is related to the increase of the effective fermion mass $m_F$ at the Fermi level and the corresponding peak in the density of states.

The role of electron-electron interactions in the behaviour of two-dimensional (2D) electron systems increases as the electron density is decreased. The interaction strength is characterized by the Wigner-Seitz radius, $r_s=1/(\pi n_s)^{1/2}a_B$ (here $n_s$ is the electron density and $a_B$ is the effective Bohr radius in semiconductor), which in the single-valley case is equal to the ratio of the Coulomb and kinetic energies. At high electron densities, where $r_s\sim 1$, interactions are not prevalent; 2D electron systems exhibit metallic conductivity and their behaviour can be well described by the Fermi liquid theory. In the opposite limit of low electron densities ($r_s>35$) interactions are dominant, and the electrons are expected to form a Wigner crystal.\cite{chaplik72,tanatar89} At intermediate values of $r_s$, where the energy of interactions exceeds all other energy scales but is still not high enough to cause the electrons to crystallize, different scenarios of the system behaviour were suggested theoretically, and numerous experiments on different electron systems were performed (see reviews \cite{kravchenko04,shashkin05,pudalov06,spivak10} and references therein). Nevertheless, the behaviour of the strongly interacting 2D electron system is still not well understood. One of the important points is the residual scattering potential. Even in quite perfect electron systems the electron properties are influenced by the residual disorder.\cite{vitkalov02,vakili04,pudalov02} At present, it is unclear whether the clean limit ({\it i.e.}, the limit in which the influence of disorder on the electron properties is negligible) has been reached in experiments on the least disordered electron systems. Finally, there exist experimental facts that contradict both intuitive expectations and calculations, {\it e.g.}, a decrease of the effective mass in fully spin-polarized single-valley electron systems.\cite{padmanabhan08,gokmen09} In view of these discrepancies it becomes imperative to obtain new experimental data on the behaviour of strongly interacting 2D electron systems as close as possible to the ideal ones in the sense of both strength of electron-electron interactions and weakness of the scattering potential.

Raw experimental data obtained in strongly correlated 2D electron systems can be divided into two groups: (i) data describing the electron system as a whole, like the magnetic field required to fully polarize electron spins, thermodynamic density of states, or magnetization of the electron system, and (ii) data related solely to the electrons at the Fermi level, like the amplitude of the Shubnikov-de~Haas oscillations yielding the effective mass $m_F$ and Lande $g$-factor $g_F$ at the Fermi level. As a rule, the data in the first group are interpreted using the quasiparticle language in which the energy-averaged values of effective mass, $m$, and Lande $g$-factor, $g$, are used. To determine the values, the formulas that hold for the case of non-interacting electrons are employed. Although this approach is ideologically incorrect, the results for $m$ and $g$ often turn out to be the same as the results for $m_F$ and $g_F$. Particularly, simultaneous increase of the energy-averaged effective mass and that at the Fermi level was reported in earlier publications.\cite{kravchenko04,shashkin05,pudalov06,dolgopolov15,mokashi12} The most pronounced effects were observed in the 2D electron system in silicon metal-oxide-semiconductor field-effect transistors (MOSFETs) in which the effective mass is strongly enhanced at low densities while the $g$-factor stays close to its value in bulk silicon, the exchange effects being small.\cite{kravchenko04,shashkin05} The strongly enhanced effective mass in Si MOSFETs was previously interpreted in favor of the occurrence of Wigner crystal or an intermediate phase (like ferromagnetic liquid). In fact, the origin and presence of possible intermediate phases preceding the formation of Wigner crystal can depend on the degree of disorder in the electron system. Because the mobility in SiGe/Si/SiGe quantum wells is two orders of magnitude higher compared to Si MOSFETs, the origin of the low-density phases in these electron systems can be different. Note that the experimental results obtained in the least-disordered Si MOSFETs exclude the metal-insulator transition driven by localization. The disorder effects in higher mobility SiGe/Si/SiGe quantum wells should be yet smaller.

\begin{figure}[ht]
\centering
\includegraphics[width=0.7\linewidth]{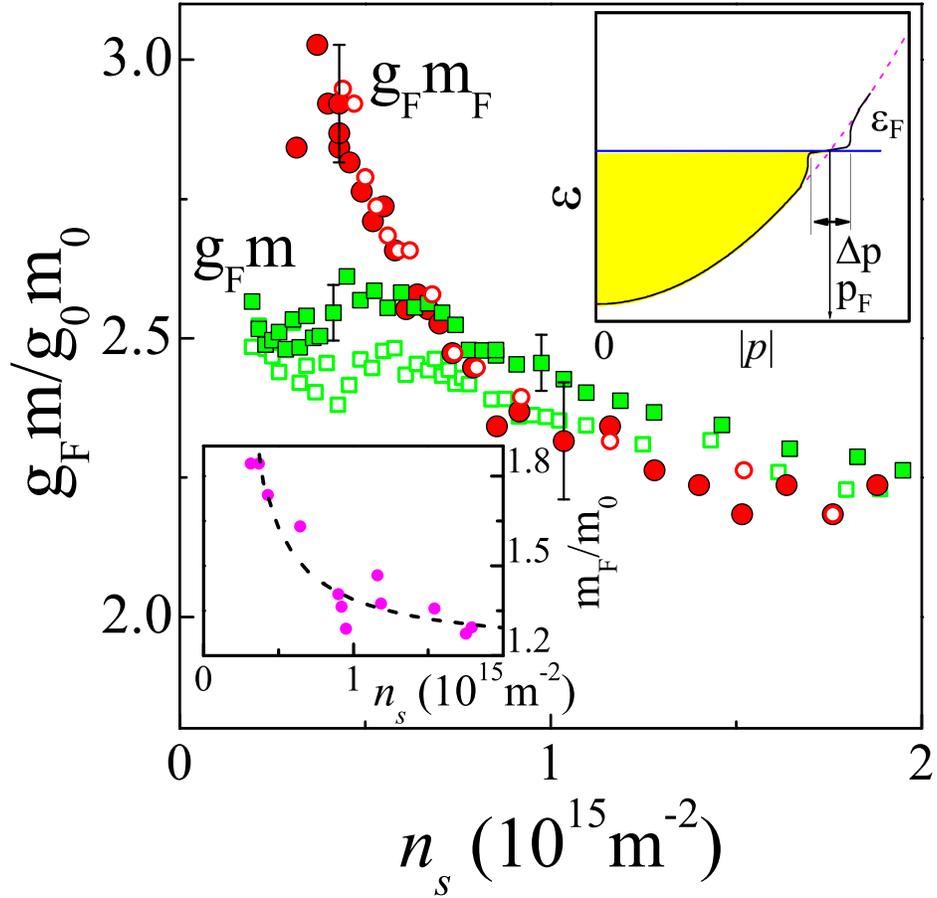}
\caption{Product of the Lande factor and effective mass as a function of electron density determined by measurements of the field of full spin polarization, $B_c$, (squares) and Shubnikov-de Haas oscillations (circles) at $T\approx 30$~mK. The empty and filled symbols correspond to two samples. The experimental uncertainty corresponds to the data dispersion and is about 2\% for the squares and about 4\% for the circles. ($g_0=2$ and $m_0=0.19m_e$ are the values for noninteracting electrons). The top inset shows schematically the single-particle spectrum of the electron system in a state preceding the band flattening at the Fermi level (solid black line). The dashed violet line corresponds to an ordinary parabolic spectrum. The occupied electron states at $T=0$ are indicated by the shaded area. Bottom inset: the effective mass $m_F$ versus electron density determined by analysis of the temperature dependence of the amplitude of Shubnikov-de Haas oscillations, similar to Ref.~\cite{melnikov14}. The dashed line is a guide to the eye.}
\label{fig1}
\end{figure}

In this work we have investigated the region of complete polarization of the electron spins in the ultra-high-mobility 2D electron system in SiGe/Si/SiGe quantum wells at millikelvin temperatures, yielding the energy-averaged product $g_Fm$, and Shubnikov-de~Haas oscillations, yielding the product $g_Fm_F$ at the Fermi level (Fig.~\ref{fig1}). We find that with decreasing electron density (or increasing interaction strength), the product $g_Fm_F$ at the Fermi level monotonically increases in the entire range of electron densities, while the energy-averaged product $g_Fm$ saturates at low densities. Taking into account the negligibility of the exchange effects in the 2D electron system in silicon, this difference can only be attributed to the different behaviours of the two effective masses. Their qualitatively different behaviour reveals a precursor to the interaction-induced single-particle spectrum flattening at the Fermi level in this electron system.

\begin{figure}[ht]
\centering
\includegraphics[width=0.7\linewidth]{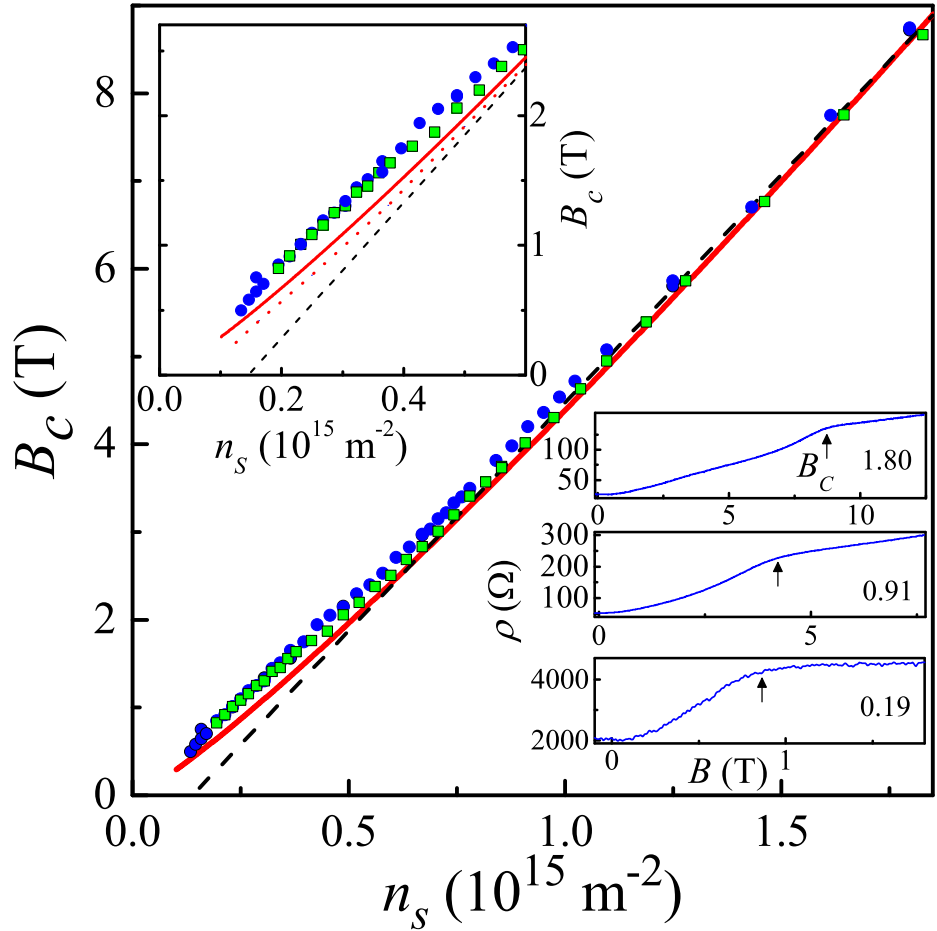}
\caption{Dependence of the field of complete spin polarization, $B_c$, on electron density at a temperature of 30~mK for two samples (dots and squares). The dashed black line is a linear fit to the high-density data which extrapolates to zero at a density $n_c$. The solid red line corresponds to the calculation \cite{fleury10} for the clean limit. Top inset: the low density region of the main figure on an expanded scale. Also shown by the dotted red line is the calculation \cite{fleury10} taking into account the electron scattering. Bottom inset: the parallel-field magnetoresistance at a temperature of 30~mK at different electron densities indicated in units of $10^{15}$~m$^{-2}$. The polarization field $B_c$ determined by the crossing of the tangents is marked by arrows.}
\label{fig2}
\end{figure}

The parallel-field magnetoresistance ({\i.e.}, magnetoresistance measured in the configuration where the magnetic field is parallel to the 2D plane) has been studied in the perpendicular orientation of the magnetic field and the current. In this case, the shape of the experimental dependences is the closest to that expected theoretically;\cite{dolgopolov00} the detailed data taken in both perpendicular and parallel orientations of the current with respect to the in-plane magnetic field will be published elsewhere. The field of the full spin polarization, $B_c$, corresponds to a distinct ``knee'' of the experimental dependences followed by the saturation of the resistance (see the bottom inset to Fig.~\ref{fig2}). The magnetic field where the spin polarization becomes complete is plotted as a function of electron density, $B_c(n_s)$, in Fig.~\ref{fig2} for two samples. Over the electron density range $0.7\times 10^{15}$~m$^{-2}<n_s<2\times 10^{15}$~m$^{-2}$, the data are described well by a linear dependence that extrapolates to zero at $n_c\approx 0.14\times 10^{15}$~m$^{-2}$ (dashed black line). However, at lower electron densities down to $n_s\approx 0.2\times 10^{15}$~m$^{-2}$ (up to $r_s\approx 12$), the experimental dependence $B_c(n_s)$ deviates from the straight line, and it linearly extrapolates to the origin. Note that the observed behaviour of the polarization field $B_c$ is quite opposite to that in the presence of the orbital effects due to the finite thickness of a 2D electron system, which are known to lead to a downward deviation in the dependence $B_c(n_s)$ with increasing electron density.\cite{zhu03,tutuc03} Therefore, the observed behaviour of $B_c$ is not related to the finite thickness effects.

The solid red line in Fig.~\ref{fig2} shows the polarization field $B_c(n_s)$ calculated using the quantum Monte Carlo method.\cite{fleury10} The experimental results are in good agreement with the theoretical calculations for the clean limit $k_Fl\gg 1$ (here $k_F$ is the Fermi wavevector and $l$ is the mean free path), assuming that the Lande $g$-factor, renormalized by electron-electron interactions, is equal to 2.4. Although in Ref.~\cite{fleury10} Lande $g$-factor was equal to 2, the reason for the 20\% discrepancy between the theory and experiment may be due to the finite size of the electron wave function in the direction perpendicular to the interface. Besides, the product $k_Fl$ decreases with decreasing electron density, which leads to a downward deviation in the theoretical dependence, as shown by the dotted red line in the upper inset to Fig.~\ref{fig2}.

To check whether or not the residual disorder affects the results for the magnetic field of complete spin polarization, we compare our data with those previously obtained on samples with an order of magnitude lower mobility.\cite{lu08} At high electron densities, the dependence $B_c(n_s)$ in Ref.~\cite{lu08} is also linear and extrapolates to zero at a finite density; the slope of the dependence is equal to $6\times 10^{-15}$~T$\cdot$m$^2$ and is close to the slope $5.4\times 10^{-15}$~T$\cdot$m$^2$ observed in our experiment. However, the offset equal approximately to $0.3\times 10^{15}$~m$^{-2}$ of Ref.~\cite{lu08} is appreciably higher compared to our case. Therefore, the behaviour of the polarization field $B_c$ is affected by the disorder potential in accordance with Refs.~\cite{fleury10,renard15}. Good agreement between our experimental data for $B_c$ and the calculations for the clean limit \cite{fleury10} gives evidence that the electron properties studied in our samples are only weakly sensitive to the residual disorder and the clean limit has been reached in our samples.

The product $g_Fm$ that characterizes the whole 2D electron system can be determined in the clean limit from the equality of the Zeeman splitting and the Fermi energy of the spin-polarized electron system
\begin{equation}
g_F\mu_BB_c=\frac{2\pi\hbar^2n_s}{mg_v},\label{gm}
\end{equation}
where $g_v=2$ is the valley degeneracy and $\mu_B$ is the Bohr magneton.

\begin{figure}[ht]
\centering
\includegraphics[width=0.7\linewidth]{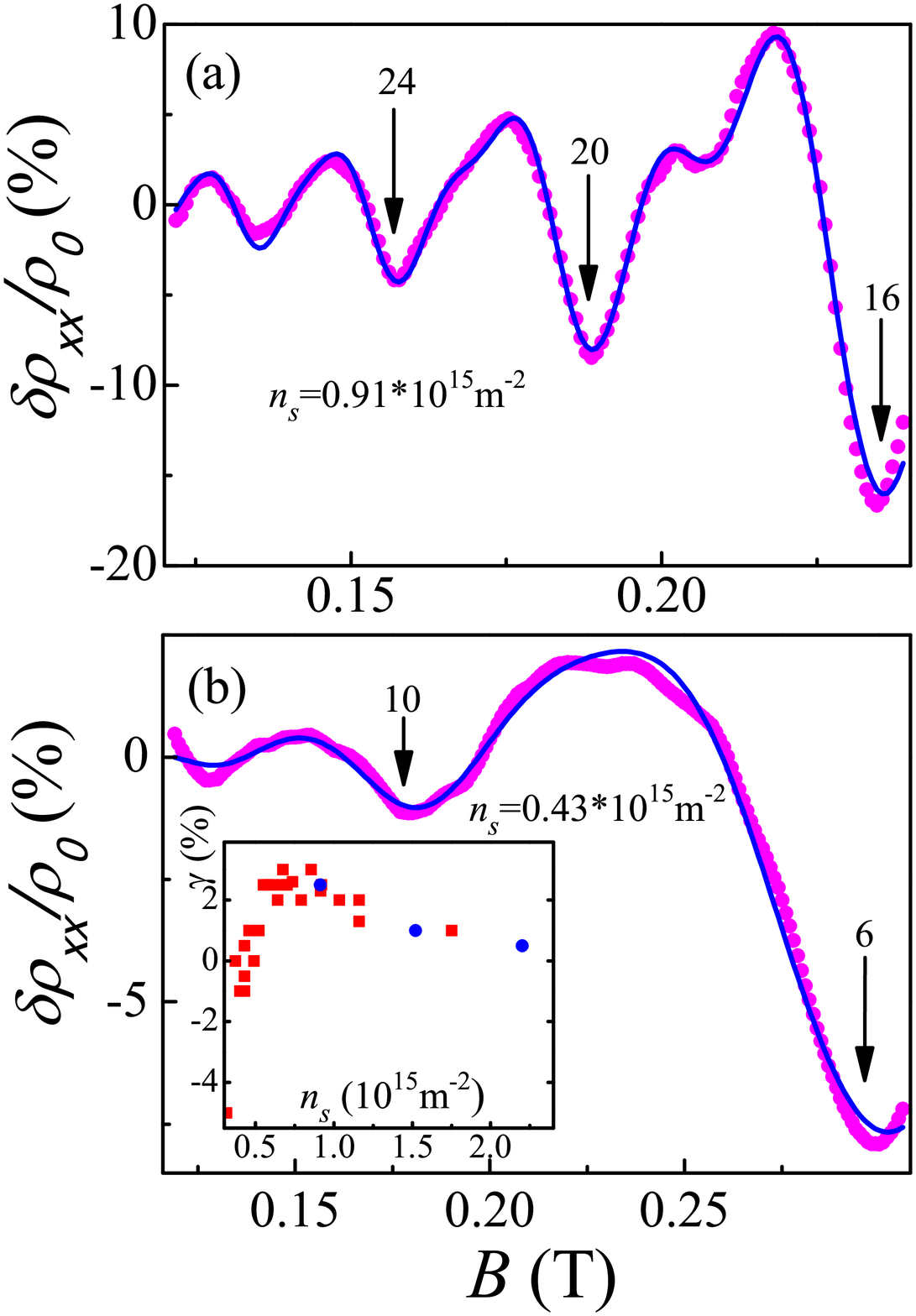}
\caption{Fits of the normalized magnetoresistance $\delta\rho_{xx}/\rho_0$ at a temperature of $\approx 30$~mK (dots) using Eq.~(\ref{A}) with (a) $g_Fm_F/m_e=0.905$, $T_D=0.12$~K, $m_F=0.25m_e$, and $\gamma=2.5$\% and (b) $g_Fm_F/m_e=1.11$, $T_D=0.15$~K, $m_F=0.33m_e$, and $\gamma=-0.5$\%. The filling factors $\nu=n_shc/eB_\perp$ at minima are indicated. Inset: the asymmetry coefficient $\gamma$ versus electron density for two samples. The Dingle temperature for two spin subbands is found to be different in our samples, similar to the results for Si MOSFETs of Ref.~\cite{pudalov14}. Although the effect is appreciably weaker in our case, we have to introduce another fitting parameter $\gamma$ for $T^{u,d}_D=T_D(1\pm\gamma)$. The difference between the Dingle temperatures for two spin subbands does not exceed 6\%, the Dingle temperature for energetically favorable spin direction being smaller over the range of electron densities $0.6\times 10^{15}$~m$^{-2}<n_s<2\times 10^{15}$~m$^{-2}$, whereas at lower densities the quantity $\gamma$ changing sign.}
\label{fig3}
\end{figure}

On the other hand, the Lande $g$-factor $g_F$ and effective mass $m_F$ \textit{at the Fermi level} can be determined by the analysis of the Shubnikov-de Haas oscillations in relatively weak magnetic fields
\begin{eqnarray}
A&=&\sum_iA^{LK}_i\cos\left[\pi i\left(\frac{\hbar c\pi n_s}{eB_\perp}-1\right)\right]Z^s_iZ^v_i\nonumber\\
A^{LK}_i&=&4\exp\left(-\frac{2\pi^2ik_BT_D}{\hbar\omega_c}\right)\frac{2\pi^2ik_BT/\hbar\omega_c}{\sinh\left(2\pi^2ik_BT/\hbar\omega_c\right)}\nonumber\\
Z^s_i&=&\cos\left(\pi i\frac{\Delta_Z}{\hbar\omega_c}\right)=\cos\left(\pi i\frac{g_Fm_F}{2m_e}\right)\nonumber\\
Z^v_i&=&\cos\left(\pi i\frac{\Delta_v}{\hbar\omega_c}\right),\label{A}
\end{eqnarray}
where $T_D$ is the Dingle temperature, $T$ is the temperature, $m_e$ is the free electron mass, $\hbar\omega_c$ is the cyclotron splitting, $\Delta_Z$ is the Zeeman splitting, and $\Delta_v$ is the valley splitting. It is clear from the relation (\ref{A}) that as long as one sets $Z^v_i=1$ in the range of magnetic fields studied, the fitting parameters are $T_Dm_F$, $m_F$, and $g_Fm_F$.\cite{pudalov14} The values $T_Dm_F$ and $m_F$ are obtained in the temperature range where the spin splitting is insignificant; the detailed results for the effective mass will be published elsewhere. Being weakly sensitive to these two fitting parameters, the shape of the fits at the lowest temperatures turns out to be very sensitive to the product $g_Fm_F$. The quality of the fits is demonstrated in Fig.~\ref{fig3}. The magnetoresistance $\delta\rho_{xx}=\rho_{xx}-\rho_0$ normalized by $\rho_0$ (where $\rho_0$ is the monotonic change of the dissipative resistivity with magnetic field) is described well using Eq.~(\ref{A}).

The main result of our study shown in Fig.~\ref{fig1} is that the products of the average $g_Fm$ and $g_Fm_F$ at the Fermi level behave similarly at high electron densities, where electron-electron interactions are relatively weak, but differ at low densities, where the interactions become especially strong. The product $g_Fm_F$ monotonically increases as the electron density is decreased in the entire range of electron densities, while the product $g_Fm$ saturates at low $n_s$. We emphasize that it is the qualitative difference in the behaviours of the two sets of data that matters, rather than comparison of the absolute values. Taking into account the negligibility of the exchange effects in the 2D electron system in silicon,\cite{kravchenko04,shashkin05} this difference can only be attributed to the different behaviours of the two effective masses. Their qualitatively different behaviour indicates the interaction-induced band flattening at the Fermi level in this electron system. To add confidence in our results and conclusions, we show in bottom inset in Fig.~\ref{fig1} the data for the effective mass $m_F$ determined by the analysis of the temperature dependence of the amplitude of Shubnikov-de Haas oscillations. The similar behaviour of $m_F$ and $g_Fm_F$ with electron density allows one to exclude any possible influence of the $g$-factor on the behaviour of the product of the effective mass and $g$-factor, which is consistent with the previously obtained results for the 2D electron system in silicon.

The experimental results are naturally interpreted within the concept of the fermion condensation \cite{khodel90,nozieres92,zverev12} that occurs at the Fermi level in a range of momenta, unlike the condensation of bosons. With increasing strength of electron-electron interactions, the single-particle spectrum flattens in a region $\Delta p$ near the Fermi momentum $p_F$ (top inset to Fig.~\ref{fig1}). At relatively high electron densities $n_s>0.7\times 10^{15}$~m$^{-2}$, this effect is not important since the single-particle spectrum does not change noticeably in the interval $\Delta p$ and the behaviours of the energy-averaged effective mass and that at the Fermi level are practically the same. Decreasing the electron density in the range $n_s<0.7\times 10^{15}$~m$^{-2}$ gives rise to the flattening of the spectrum so that the effective mass at the Fermi level, $m_F=p_F/V_F$, continues to increase (here $V_F$ is the Fermi velocity). In contrast, the energy-averaged effective mass does not, being not particularly sensitive to this flattening. Near the density $n_c$, the electron system is in a critical region in which the effective mass at the Fermi level is expected to be temperature dependent. A weak decrease of the value $g_Fm_F$ with temperature is indeed observed at the lowest-density point in Fig.~\ref{fig1}. In the critical region, the increase of $m_F$ is restricted by the limiting value determined by temperature: $m_F<p_F\Delta p/4k_BT$. In our experiments, the increase of $m_F$ reaches a factor of about two at $n_s=0.3\times 10^{15}$~m$^{-2}$ and $T\approx 30$~mK, which allows one to estimate the ratio $\Delta p/p_F\sim 0.06$. It is the smallness of the interval $\Delta p$ that provides good agreement between the calculation \cite{fleury10} and our experiment.

\section*{Methods}

Samples were prepared based on a 15~nm SiGe/Si/SiGe quantum well grown in an ultrahigh-vacuum chemical-vapor-deposition (UHVCVD) apparatus.\cite{lu09,lu10} The Hall-bar samples with width 50~$\mu$m and the distance between potential probes 150~$\mu$m were patterned using standard photo-lithography. Contacts consisted of AuSb alloy, deposited in a thermal evaporator in vacuum and annealed for 5 minutes in N$_2$ atmosphere at 440$^\circ$C. Next, approximately 300~nm thick layer of SiO was deposited in a thermal evaporator and a $>20$~nm thick Al gate was deposited on top of SiO. No mesa etching was used, and the 2D electron gas was created in a way similar to silicon MOSFETs (for details, see Refs.~\cite{melnikov14,melnikov15}). The maximum electron mobility in our samples reached 240~m$^2$/Vs which is the highest mobility reported for this electron system.\cite{melnikov15}

Measurements were made in an Oxford TLM-400 dilution refrigerator in a temperature range 0.03 -- 1.2~K. The resistance was measured by a standard four-terminal lock-in technique in a frequency range 1 -- 11~Hz. The applied currents varied in the range 0.5 -- 4~nA. We used a saturating infra-red illumination to improve the quality of the contacts and increase the electron mobility. This did not affect the electron density at a fixed gate voltage.

\section*{Acknowledgements}

This work was supported by RFBR 15-02-03537 and 16-02-00404, RAS, and the Russian Ministry of Sciences. NTU Group was supported by Ministry of Science and Technology (104-2622-8-002-003). S.V.K. was supported by NSF Grant No.\ 1309337 and BSF Grant No.\ 2012210.

\section*{Author contributions statement}

M.Yu.M., A.A.S. and V.T.D. prepared samples, conceived and conducted the experiments, analysed the results, developed the model, and composed the manuscript.  S.S.H. and C.W.L. grew SiGe/Si/SiGe wafers and contributed to the discussions and composition of the article.  S.V.K. contributed to the discussions and composition of the article.  All authors reviewed the manuscript.

\section*{Additional Information}

Competing Financial Interest: There is NO Competing Interest.

\pagebreak

\begin{thebibliography}{1}
\bibitem{heikkila2011} Heikkila, T.~T., Kopnin, N.~B. \& Volovik, G.~E. Flat bands in topological media. \textit{JETP\ Lett}. {\bf 94}, 233-239 (2011).
\bibitem{ns2013} Bennemann, K.-H. \& and Ketterson, J.~B. \textit{Novel Superfluids} (Oxford University Press, 2013).
\bibitem{peotta2015} Peotta, S. \& and Torma, P. Superfluidity in topologically nontrivial flat bands. \textit{Nat.\ Commun.}\ {\bf 6}, 8944 (2015).
\bibitem{volovik15} Volovik, G.~E. From standard model of particle physics to room-temperature superconductivity. \textit{Phys.\ Scr.} {\bf T164}, 014014 (2015).
\bibitem{amusia14} Amusia, M.~Ya., Popov, K.~G., Shaginyan, V.~R. \& Stephanovich, V.~A. {\it Theory of Heavy Fermion Compounds} (Springer, Berlin, 2014).
\bibitem{camjayi08} Camjayi, A., Haule, K., Dobrosavljevi\'c, V. \& Kotliar, G. Coulomb Correlations and the Wigner-Mott Transition. \textit{Nat.\ Phys}. {\bf 4}, 932-935 (2008).
\bibitem{yudin14} Yudin, D., Hirschmeier, D., Hafermann, H., Eriksson, O., Lichtenstein, A.~I. \& Katsnelson, M.~I. Fermi condensation near van Hove singularities within the Hubbard model on the triangular lattice.
\textit{Phys.\ Rev.\ Lett.} \textbf{112}, 070403 (2014).
\bibitem{chaplik72} Chaplik, A.~V. Possible crystallization of charge carriers in low-density inversion layers. {\it Sov.\ Phys.\ JETP} {\bf 35}, 395-398 (1972).
\bibitem{tanatar89} Tanatar, B. \& Ceperley, D.~M. Ground state of the two-dimensional electron gas. {\it Phys.\ Rev.\ B} {\bf 39}, 5005-5016 (1989).
\bibitem{kravchenko04} Kravchenko, S.~V. \& Sarachik, M.~P. Metal-insulator transition in two-dimensional electron systems. {\it Rep.\ Prog.\ Phys.} {\bf 67}, 1-44 (2004).
\bibitem{shashkin05} Shashkin, A.~A. Metal-insulator transitions and the effects of electron-electron interactions in two-dimensional electron systems. {\it Phys.\ Usp.} {\bf 48}, 129-149 (2005).
\bibitem{pudalov06} Pudalov, V.~M. Metal-insulator transitions and related phenomena in a strongly correlated two-dimensional electron system. {\it Phys.\ Usp.} {\bf 49}, 203-208 (2006).
\bibitem{spivak10} Spivak, B., Kravchenko, S.~V., Kivelson, S.~A. \& Gao, X.~P.~A. Transport in strongly correlated two dimensional electron fluids. \textit{Rev.\ Mod.\ Phys.}\ {\bf 82}, 1743-1766 (2010).
\bibitem{vitkalov02} Vitkalov, S.~A., Sarachik, M.~P. \& Klapwijk, T.~M. Spin polarization of strongly interacting two-dimensional electrons: The role of disorder. {\it Phys.\ Rev.\ B} {\bf 65}, 201106(R) (2002).
\bibitem{vakili04} Vakili, K., Shkolnikov, Y.~P., Tutuc, E., De Poortere, E.~P. \& Shayegan, M. Spin susceptibility of two-dimensional electrons in narrow AlAs quantum wells. {\it Phys.\ Rev.\ Lett.} {\bf 92}, 226401 (2004).
\bibitem{pudalov02} Pudalov, V.~M., Brunthaler, G., Prinz, A., \& Bauer, G. Weak anisotropy and disorder dependence of the in-plane magnetoresistance in high-mobility (100) Si-inversion layers. {\it Phys.\ Rev.\ Lett.} {\bf 88}, 076401 (2002).
\bibitem{padmanabhan08} Padmanabhan, M., Gokmen, T., Bishop, N.~C. \& Shayegan, M. Effective mass suppression in dilute, spin-polarized two-dimensional electron systems. {\it Phys.\ Rev.\ Lett.} {\bf 101}, 026402 (2008).
\bibitem{gokmen09} Gokmen, T., Padmanabhan, M., Vakili, K., Tutuc, E. \& Shayegan, M. Effective mass suppression upon complete spin-polarization in an isotropic two-dimensional electron system. {\it Phys.\ Rev.\ B} {\bf 79}, 195311 (2009).
\bibitem{dolgopolov15} Dolgopolov, V.~T. Two-dimensional electrons in (100)-oriented silicon field-effect structures in the region of low concentrations and high mobilities. {\it JETP\ Lett.} {\bf 101}, 282-287 (2015).
\bibitem{mokashi12} Mokashi, A., Li, S., Wen, B., Kravchenko, S.~V., Shashkin, A.~A., Dolgopolov, V.~T., \& Sarachik, M.~P. Critical behavior of a strongly interacting 2D electron system. {\it Phys.\ Rev.\ Lett.} {\bf 109}, 096405 (2012).
\bibitem{dolgopolov00} Dolgopolov, V.~T. \& Gold, A. Magnetoresistance of a two-dimensional electron gas in a parallel magnetic field. {\it JETP\ Lett.} {\bf 71}, 27-30 (2000).
\bibitem{zhu03} Zhu, J., Stormer, H.~L., Pfeiffer, L.~N., Baldwin, K.~W. \& West, K.~W. Spin Susceptibility of an Ultra-Low-Density Two-Dimensional Electron System. \textit{Phys.\ Rev.\ Lett.}\ {\bf 90}, 056805 (2003).
\bibitem{tutuc03} Tutuc, E., Melinte, S., De~Poortere, E.~P., Shayegan, M. \& Winkler, R. Role of finite layer thickness in spin polarization of GaAs two-dimensional electrons in strong parallel magnetic fields.  \textit{Phys.\ Rev.}\ B\ {\bf 67}, 241309(R) (2003).
\bibitem{fleury10} Fleury, G. \& Waintal, X. Energy scale behind the metallic behaviors in low-density Si MOSFETs. {\it Phys.\ Rev.\ B} {\bf 81}, 165117 (2010).
\bibitem{lu08} Lu, T.~M., Sun, L., Tsui, D.~C., Lyon, S., Pan, W., Muehlberger, M., Shaeffler, F., Liu, J. \& Xie, Y.~H. In-plane field magnetoresistivity of Si two-dimensional electron gas in Si/SiGe quantum wells at 20 mK. {\it Phys.\ Rev.\ B} {\bf 78}, 233309 (2008).
\bibitem{renard15} Renard, V.~T., Piot, B.~A., Waintal, X., Fleury, G., Cooper, D., Niida, Y., Tregurtha, D., Fujiwara, A., Hirayama, Y. \& Takashina, K. Valley polarization assisted spin polarization in two dimensions. \textit{Nat.\ Commun.}\ {\bf 6}, 7230 (2015).
\bibitem{pudalov14} Pudalov, V.~M., Gershenson, M.~E. \& Kojima, H. Probing electron interactions in a two-dimensional system by quantum magneto-oscillations. {\it Phys.\ Rev.\ B} {\bf 90}, 075147 (2014).
\bibitem{khodel90} Khodel, V.~A. \& Shaginyan, V.~R. Superfluidity in system with fermion condensate. {\it JETP\ Lett.} {\bf 51}, 553-555 (1990).
\bibitem{nozieres92} Nozi\`eres, P.\ Properties of Fermi liquids with a finite range interaction. \textit{J.\ Phys.\ I (France)} {\bf 2}, 443-458 (1992).
\bibitem{zverev12} Zverev, M.~V., Khodel, V.~A. \& Pankratov, S.~S. Microscopic theory of a strongly correlated two-dimensional electron gas. {\it JETP\ Lett.} {\bf 96}, 192-202 (2012).
\bibitem{lu09} Lu, T.~M., Tsui, D.~C., Lee, C.-H. \& Liu, C.~W. Observation of two-dimensional electron gas in a Si quantum well with mobility of $1.6\times10^6$ cm$^2$/Vs. {\it Appl.\ Phys.\ Lett.} {\bf 94}, 182102 (2009).
\bibitem{lu10} Lu, T.~M., Tsui, D.~C., Lee, C.-H. \& Liu, C.~W. Erratum: Observation of two-dimensional electron gas in a Si quantum well with mobility of $1.6\times10^6$ cm$^2$/Vs [\textit{Appl. Phys. Lett}. \textbf{94}, 182102 (2009)]. {\it Appl.\ Phys.\ Lett.} {\bf 97}, 059904 (2010).
\bibitem{melnikov14} Melnikov, M.~Yu., Shashkin, A.~A., Dolgopolov, V.~T., Kravchenko, S.~V., Huan, S.-H. \& Liu, C.~W. Effective electron mass in high-mobility SiGe/Si/SiGe quantum wells. {\it JETP\ Lett.} {\bf 100}, 114-119 (2014).
\bibitem{melnikov15} Melnikov, M.~Yu., Shashkin, A.~A., Dolgopolov, V.~T., Huan, S.-H., Liu, C.~W. \& Kravchenko, S.~V. Ultra-high mobility two-dimensional electron gas in a SiGe/Si/SiGe quantum well. {\it Appl.\ Phys.\ Lett.} {\bf 106}, 092102 (2015).
\end{thebibliography}
\end{document}